\documentclass[floatfix,aps,prd,showpacs,amsmath,nofootinbib,preprintnumbers,twocolumn]{revtex4}

\usepackage{graphicx}
\usepackage{bm}

\newcommand{\figurewidth}{3.4in}

\def\e{\begin{equation}}
\def\q{\end{equation}}
\def\m{\begin{eqnarray}}
\def\n{\end{eqnarray}}

\def\half{{1\over 2}}
\def\p{\partial}

\def\[{\left [}
\def\]{\right ]}
\def\({\left (}
\def\){\right )}

\begin{document}

\title{Constraints on the spectral index for the inflation models in string landscape}

\author{Qing-Guo Huang}\email{huangqg@kias.re.kr}
\affiliation{School of physics, Korea Institute for Advanced Study,
207-43, Cheongryangri-Dong, Dongdaemun-Gu, Seoul 130-722, Korea}

\date{\today}

\begin{abstract}

We conjecture that the inflation models with trans-Planckian
excursions in the field space should be in the swampland. We check
this conjecture in a few examples and investigate the constraints on
the spectral index for the slow-roll inflation model in string
landscape where the variation of inflaton during the period of
inflation is less than the Planck scale $M_p$. A red primordial
power spectrum with a lower bound on the spectral index is
preferred. Both the tensor-scalar ratio and the running can be
ignored.

\end{abstract}


\maketitle


Inflation model \cite{Guth:1980zm,Linde:1981mu,Albrecht:1982wi} has
been remarkably successful in not only explaining the large-scale
homogeneity and isotropy of the universe, but also providing a
natural mechanism to generate the observed magnitude of
inhomogeneity. In the new version of inflation, inflation may begin
either in the false vacuum, or in an unstable state at the top of
the effective potential. Then the inflaton field slowly rolls down
to the minimum of its effective potential. This picture relies on an
application of low-energy effective field theory to inflation.
However, the effective field theory can break down even in the
region with low curvature.

It is clear that consistent theories of quantum gravity can be
constructed in the context of string theory. The central problem in
string theory is how to connect it with experiments. Recent
developments for the flux compactifications
\cite{Giddings:2001yu,Kachru:2003aw} suggest that a vast number of
at least semi-classically consistent string vacua emerge in string
theory. It may or may not provide an opportunity for us to explore
the specific low energy phenomena in the experiments from the
viewpoint of string theory. In fact the vast series of
semi-classically consistent effective field theories are actually
inconsistent. We say that they are in the swampland
\cite{Vafa:2005ui}. Self-consistent landscape is surrounded by the
swampland. In \cite{Arkani-Hamed:2006dz} gravity is conjectured as
the weakest force on the validity of the effective field theories.
This conjecture is supported by string theory and some evidence
involving black holes and symmetries. In four dimensions an
intrinsic UV cutoff for the U(1) gauge theory \e \Lambda\leq gM_p
\label{wggt}\q is suggested, where $g$ is the gauge coupling and
$M_p$ is the Planck scale. Furthermore, an intrinsic UV cutoff for
the scalar field theories with gravity is proposed in
\cite{Huang:2007gk}, e.g. \e \Lambda\leq \lambda^{1/2}M_p
\label{wgcs}\q for $\lambda\phi^4$ theory. This conjecture provides
some possible stringent constraints on inflation model
\cite{Huang:2007gk}. Some other related works on weak gravity
conjecture are
\cite{Kachru:2006em,Li:2006vc,Huang:2006hc,Li:2006jj,Kats:2006xp,Banks:2006mm,Huang:2006pn,Medved:2006ht,Huang:2006tz,Huang:2007mf,Eguchi:2007iw}.

The gauge interactions are governed by the symmetry. However there
is not such a principle to constrain the interaction of scalar
fields. Ones can construct thousands of inflation models
corresponding to different shapes of the potential of inflaton.
Therefore it is difficult for us to work out some model-independent
predictions for inflation model.

In this note, we collect several examples to support that the
variation of the inflaton for the inflation models in string
landscape should be less than the Planck scale. According to this
observation, we figure out the constraints on the spectral index for
the inflation model.

Inflation in the early universe is driven by the potential of the
inflaton field $\phi$. The equations of motion for an expanding
spatially flat universe containing a homogeneous scalar field take
the form \m H^2=\({\dot a \over a}\)^2&=&{1\over 3M_p^2}\(\half
{\dot \phi}^2+V(\phi)\),
\\ \ddot \phi+3H\dot \phi&=&-V', \n where $V(\phi)$ is the potential
of inflaton $\phi$ and the prime denotes the derivative with respect
to $\phi$. For simplicity, we define several slow-roll parameters as
\e \epsilon={M_p^2\over 2}\({V'\over V}\)^2, \eta=M_p^2{V''\over V},
\xi=M_p^4{V'V'''\over V^2}.\q If $\epsilon\ll 1$ and $|\eta|\ll 1$
the inflaton field slowly rolls down its potential and the equations
of motion are simplified to be \e H^2={V\over 3M_p^2}, \quad 3H\dot
\phi=-V'.\q In this paper, we assume, without loss of generality,
$\dot \phi<0$, so that $V'>0$. The number of e-folds $N$ before the
end of inflation is related to the vev (vacuum expectation value) of
inflaton by \e dN=-Hdt=-{H\over \dot \phi}d\phi={1\over
\sqrt{2\epsilon}M_p}d\phi. \label{np}\q The slow-roll parameters
also characterize the feature of the primordial power spectrum for
the fluctuations: the amplitude of the scalar and tensor
perturbations are respectively \cite{Liddle:2000cg} \e \Delta_{\cal
R}^2={H^2/M_p^2\over 8\pi^2 \epsilon},\quad
\Delta_T^2={H^2/M_p^2\over \pi^2/2}. \q The tensor-scalar ratio
takes the form \e r=\Delta_T^2/\Delta_{\cal R}^2=16\epsilon, \q and
the spectral index and its running are given by \m
n_s&=&1-6\epsilon+2\eta,
\\ \alpha_s&=& -24\epsilon^2+16\epsilon\eta-2\xi, \n where we use \e
{d\epsilon \over dN}=2\epsilon(\eta-2\epsilon), \quad {d\eta\over
dN}=\xi-2\epsilon\eta. \label{des}\q

In \cite{Lyth:1996im} Lyth connects detectably large gravitational
wave signals to the motion of the inflaton over Planckian distances
in the field space. There is a long-term debate
\cite{Lyth:1996im,Linde:2007fr} on whether the classical evolution
of the scalar field can probe the trans-Planckian region where the
low energy field theory is still an effective field theory. String
theory gives us an opportunity to answer this question. In this
note, we conjecture that the probing region of the scalar field is
limited by the Planck scale $M_p$ in the string landscape. A few
examples to check our conjecture will be proposed as follows.

The first one is called ``extra-natural inflation"
\cite{Arkani-Hamed:2003wu}. Consider a U(1) gauge theory with gauge
coupling $g_5$ in five dimensions. Compactifying this gauge theory
on a circle with size $R$, we obtain four-dimensional gravity as
well as a periodic scalar $\theta=\oint A_5 dx^5$ associated with
the Wilson line around the circle. The effective Lagrangian for
$\theta$ in four dimensions takes the form \e {\cal L}={f^2\over
2}(\p \theta)^2-{1\over R^4}(1-\cos \theta), \q where $f^2={1\over
g_5^2R}={1\over g^2R^2}$ and $g$ is the gauge coupling in four
dimensions. The canonical scalar field $\phi$ is given by
$\phi=f\theta$. The period of $\theta$ is $2\pi$ and the vev of
$\phi$ takes the same order of magnitude as $f$. It is easily seen
that $f$ can be bigger than $M_p$ for sufficiently small $g$ and the
slow-roll conditions are achieved. However the weak gravity
conjecture \cite{Arkani-Hamed:2006dz} says $\Lambda\sim 1/R \leq
gM_p$ which implies $f={1\over gR}\leq M_p$. With the viewpoint of
string theory, $g=g_s^{1/2}/\sqrt{M_s^6V_6}$ and
$M_p=M_s\sqrt{M_s^6V_6}/g_s$, where $M_s$ is the string scale and
$V_6$ is the volume of the compactified space. Thus we have
$f={1\over gR}={g_s^{1/2}\over M_sR}M_p<M_p$ in the perturbative
region ($g_s<1$), where we also require that the size of the
compactified space is larger than string length $M_s^{-1}$. In this
case, the over Planckian excursion of the scalar field cannot be
embedded into string theory and it is in the swampland.

The second is chaotic inflation \cite{Linde:1983gd}. For an
instance, we consider $V(\phi)=\lambda \phi^4$ inflation model. The
Hubble scale $H=\sqrt{V\over 3M_p^2}\sim {\lambda^{1/2}\phi^2\over
M_p}$ can be taken as the IR cutoff for the effective field theory.
In \cite{Huang:2007gk} an upper bound on the UV cutoff (\ref{wgcs})
is proposed. Naturally the IR cutoff should be lower than the UV
cutoff. Requiring $H<\Lambda$ yields $\phi<M_p$ \cite{Huang:2007gk}.
Furthermore, we take into account the inflation model with potential
$V=V_0+\lambda \phi^4$. If the potential is dominated by the
constant term $V_0$, it is a typical potential for hybrid inflation
\cite{Linde:1993cn}. Since $H=\sqrt{V_0+\lambda\phi^4\over
3M_p^2}>{\lambda^{1/2}\phi^2\over M_p}$, requiring $H\leq \Lambda$
leads to $\phi<M_p$ as well. The trans-Planckian excursion in the
field space cannot be achieved.

The third example is the inflation driven by the motion of a
D3-brane in the warped background. The authors in
\cite{Baumann:2006cd} found the maximal variation of the canonical
inflaton field as \e |\Delta \phi|=\sqrt{T_3}R\leq {2\over
\sqrt{n_B}}M_p, \label{cbp}\q where $R$ is the size of the throat
and $n_B$ is the number of the background D3 charge. Since $n_B\gg
1$ for the validity of the background geometry, the variation of the
inflaton is not larger than the Planck scale.

Fourth Ooguri and Vafa in \cite{Ooguri:2006in} propose several
conjectures to limit the observable regions of moduli spaces. For a
massless scalar field $\phi$, the change of its vev is $|\Delta
\phi|\sim |{M_p\over 3} \ln \varepsilon|$, where $\varepsilon$ is
the mass scale for the low energy effective theory. There is an
infinite tower of light particles at infinite distance from any
point inside the moduli space, the effective field theory in the
interior breaks down and a new description takes over. This example
also hints that the variation of the scalar field should be less
than $M_p$ in string landscape.

In the following we will investigate the constraints on the spectral
index by considering that the variation of the inflaton during the
period of inflation is less than $M_p$. We re-parameterize the
slow-roll parameter $\epsilon$ in eq. (\ref{np}) as a function of
$N$. Eq. (\ref{np}) becomes \e
\int_{0}^{N_{tot}}\sqrt{2\epsilon(N)}dN=\int {d\phi\over
M_p}={|\Delta \phi|\over M_p}\leq 1. \label{cn} \q We cannot really
achieve a model-independent analysis, because the function
$\epsilon(N)$ for the string landscape is unknown. Here we consider
three typical parameterizations. Actually these parameterizations
are quite general and many well-known inflation models are included
in them.

First we assume $\epsilon$ is roughly a constant and then
$\eta=2\epsilon$. Eq. (\ref{cn}) reads \e \epsilon\leq
\epsilon_m={1\over 2N_{tot}^2}. \label{cep}\q Now the spectral index
and the tensor-scalar ratio are \e n_s=1-2\epsilon\geq 1-{1\over
N_{tot}^2}, \quad r=16\epsilon\leq {8\over N_{tot}^2}. \q
Generically the total number of e-folds should be larger than 60 in
order to solve the flatness and horizon problem. In this case, the
scalar power spectrum is the scale-invariant
HarrisonZel'dovich-Peebles (HZ) spcetrum with ignoring tensor
perturbations $r\leq 0.002$. WMAP normalization is $\Delta_{\cal
R}^2=2\times 10^9$ \cite{Spergel:2006hy}. Thus $\Delta_T^2=r\cdot
\Delta_{\cal R}^2\leq 4\times 10^{-12}$ and $V^{1/4}\leq 6.8\times
10^{15}$GeV which is lower than the GUT scale.

Second we consider the case with \e \epsilon(N)={c^2/2\over
N^{2-2\beta}}, \label{sep}\q where both $c$ and $\beta$ are
constants. Since $\epsilon<1$ for $N=60$, it is reasonable to assume
that the value of $\beta$ is not larger than 1. Requiring that the
integration in the left hand side of eq. (\ref{cn}) is finite yields
$\beta> 0$. \footnote{For example, Brane inflation
\cite{Dvali:1998pa} (KKLMMT model \cite{Kachru:2003sx}) takes
$\beta=1/6$.} Therefore the reasonable range for $\beta$ is \e 0<
\beta\leq 1.
\q Using eq. (\ref{des}) and (\ref{sep}), we obtain \m \eta&=&2\epsilon-{1-\beta\over N},\\
\xi&=&{1-\beta\over N^2}-{6(1-\beta)\over N}\epsilon+4\epsilon^2.\n
The spectral index and its running and the tensor-scalar ratio
are respectively \m n_s&=&1-2\epsilon-{2(1-\beta)\over N}, \\
\alpha_s&=&-{2(1-\beta)\over N^2}-{4(1-\beta)\over N}\epsilon, \\
r&=&16\epsilon. \n Now eq. (\ref{cn}) implies \e c\leq \beta
N_{tot}^{-\beta}.\label{cb} \q It comes back to the previous results
for $\beta=1$. Eq. (\ref{cb}) leads to an upper bound on $\epsilon$
\e \epsilon\leq {\beta^2\over 2N^2}\({N\over
N_{tot}}\)^{2\beta}.\label{ce}\q Since $0\leq \beta\leq 1$ and
$N\leq N_{tot}$, $\epsilon\leq {1\over 2N^2}=1.4\times 10^{-4}$ for
$N=60$. The tensor-scalar ratio satisfies $r\leq 0.002$. Thus
$\Delta_T^2=r\cdot \Delta_{\cal R}^2\leq 4\times 10^{-12}$ and
$V^{1/4}\leq 6.8\times 10^{15}$GeV. Since the maximum value of
$\epsilon$ takes the order of magnitude $10^{-4}$, we can ignore the
terms with $\epsilon$. Now the spectral index and its running become
\e n_s=1-{2(1-\beta)\over N}, \quad \alpha_s=-{2(1-\beta)\over N^2}
.\q Since $\beta>0$, there are the lower bounds on the spectral
index and its running: \e 1-{2\over N}\leq n_s<1,\quad -{2\over
N^2}\leq \alpha_s<0. \label{lbns}\q A red tilted primordial power
spectrum ($n_s<1$) with ignoring running and tensor perturbations
arises in string landscape. On the other hand, WMAP data
\cite{Spergel:2006hy} prefers a red tilted power spectrum: \e
n_s=0.951\pm0.016, \quad r\leq 0.65, \q and the running can be
ignored. We compare our constraints on the inflation models in the
string landscape with WMAP in fig. 1.
\begin{figure}[ht]
\centerline{\includegraphics[width=\figurewidth]{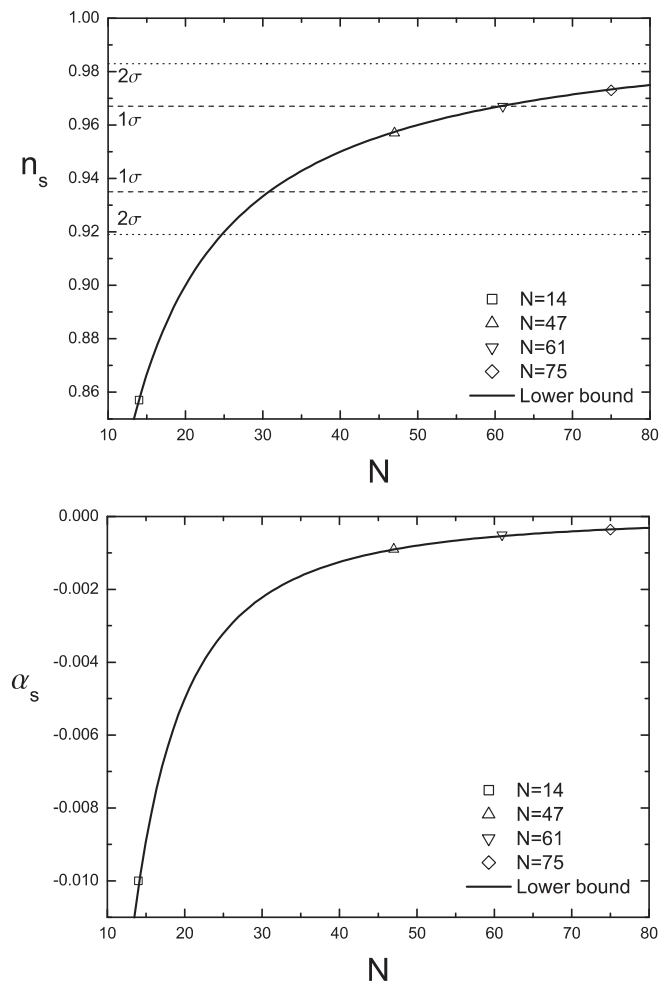}}
\caption{The lower bounds on $n_s$ and $\alpha_s$ are showed for the
single-field inflation model with sub-Planckian excursion in the
field space. With a high inflation scale, and radiation and/or
matter domination between the end of inflation and nucleosynthesis,
$47\leq N\leq 61$. More generally the range has to be $14\leq N\leq
75$ \cite{Alabidi:2006qa}.}
\end{figure}
Our analysis is consistent with observations.

Third we consider \e \epsilon(N)=\epsilon_0+{c^2/2\over
N^{2-2\beta}}, \label{tep}\q where $\epsilon_0$, $c$ and $\beta$ are
constant. Here we assume $\epsilon_0>0$ and then the range of
$\beta$ is still $\beta\in [0,1]$. In this case the constraints on
$\epsilon_0$ and $c$ should be more stringent than those in the
previous two cases, because both terms on the right hand side of eq.
(\ref{tep}) is positive. For simplicity, we still take
$\epsilon_0\leq \epsilon_m$ and $c\leq \beta N_{tot}^{-\beta}$, and
thus the terms with $\epsilon$ can be ignored. Now the slow-roll
parameters take the form \e \eta=-\alpha{1-\beta\over N}, \quad
\xi=\gamma{1-\beta\over N^2},\q where $\alpha=1/(
1+2\epsilon_0N^{2-2\beta}/c^2)\leq 1$ and
$\gamma=3-2\alpha^2-2(1-\alpha^2)\beta\leq 3$. Since
$n_s=1-2\eta=1-2\alpha{1-\beta\over N}$ and $\alpha\leq 1$, a more
blue tilted power spectrum than the previous case with
$\epsilon=c^2/(2N^{2-2\beta})$ is obtained. In this case the running
of the spectral index can be ignored as well. The lower bound on the
spectral index in eq. (\ref{lbns}) is still available.


The previous discussions are only valid for the single-field
inflation model in string landscape. For multi-field inflation, the
previous constraints may be released. To be simple, we consider the
assisted inflation \cite{Liddle:1998jc} with potential
$\sum_{i=1}^{n}V(\phi_i)$. In the assisted inflation, there is a
unique late-time attractor with all the scalar fields equal, i.e.
$\phi_1=\phi_2=...=\phi_n$. With this ansatz, the equations of
motion for the slow-roll assisted inflation are given by \e
H^2={nV(\phi)\over 3M_p^2}, \quad 3H\dot \phi=-V',\label{spv} \q
where $\phi=\phi_i, i=1,...,n$. It is convenient for us to define a
new slow-roll parameter $\epsilon_H$ as \e \epsilon_H=-{\dot H\over
H^2}. \q Slow-roll condition reads $\epsilon_H\ll 1$. Using eq.
(\ref{spv}), we find \e \epsilon_H={1\over n}{M_p^2\over
2}\({V'\over V}\)^2={1\over n}\epsilon. \q Because of the factor
$1/n$ in the above equation, the slow-roll condition for the
inflation model without flat enough potential ($\epsilon\gg 1$) can
be achieved if the number of the inflatons is sufficiently large.
Replacing $\epsilon$ in eq. (\ref{cn}) with $\epsilon_H$, we obtain
\e \int_{0}^{N_{tot}}\sqrt{2\epsilon_H(N)}dN=\sqrt{n}{|\Delta
\phi|\over M_p}. \q If we still have $|\Delta \phi|\leq M_p$ and \e
\epsilon_H(N)={c^2/2\over N^{2-2\beta}},\q the bound on $c$ becomes
\e c\leq \sqrt{n}\beta N_{tot}^{-\beta}. \q The upper bound on the
slow-roll parameter $\epsilon_H$ is given by \e \epsilon_H\leq
{n\beta^2\over 2N^2}\({N\over N_{tot}}\)^{2\beta}. \q If the number
of the inflatons $n$ is large enough, we can get a larger slow-roll
parameter $\epsilon_H$, a larger tensor-scalar ratio
$r=16\epsilon_H$ and a more red tilted power spectrum. Before the
end of this paragraph, we also want to re-consider an example in
string theory: brane inflation in the warped background. If the
number of the probing D3-branes is $n$ which is just the number of
inflatons, we have $ \sqrt{n}|\Delta \phi|\leq 2\sqrt{n\over
n_B}M_p$. In order for the validity of the background geometry,
$n<n_B$; otherwise the back reaction of the probing D3-branes will
significantly change the background geometry. In this case,
$\sqrt{n}|\Delta \phi|<M_p$. If this is the generic result for the
inflation models in the string landscape, our previous results for
the single-field inflation are recovered even for the multi-field
inflation models.

To summarize, the inflation model with over Planckian excursion in
the scalar field space cannot be achieved in string theory. A red
tilted primordial scalar power spectrum with a lower bound on the
spectral index arises for the slow-roll inflation model in string
landscape due to the observation that the observable region in the
scalar field space is limited by the Planck scale. The tensor
fluctuations and the running of the spectral index can be ignored.
Even though our analysis is not really model-independent, the
parameterizations in this note are already quite general. In some
sense, our results can be taken as the predictions of string theory.
For the assisted inflation, the constraints on the spectral index
might be released.

At last we also want to remind that maybe the chain inflation
\cite{Freese:2004vs,Feldstein:2006hm,Freese:2006fk,Huang:2007ek} is
generic in string landscape. In this model, the universe tunnelled
rapidly through a series of metastable vacua with different vacuum
energies. Since chain inflation is not really a slow-roll inflation
model, it doesn't suffer from the constraints in this paper. A
detectable gravitational wave fluctuations is still available in
this model \cite{Huang:2007ek}.

\vspace{5mm}

\noindent {\bf Acknowledgments} We would like to thank K.M. Lee,
F.L. Lin, E. Weinberg, P.J. Yi for useful discussions.



\end{document}